\documentclass[prx,twocolumn,longbibliography,superscriptaddress,preprintnumbers]{revtex4-2}
\usepackage[colorlinks,bookmarks=true,citecolor=blue,linkcolor=blue,urlcolor=blue]{hyperref}
\setcitestyle{numbers,square}
\usepackage{amsmath,amssymb,amsthm,amsfonts}
\usepackage{tabularx}
\usepackage{graphicx}
\usepackage{bmpsize}
\usepackage{bm}
\usepackage{color}
\usepackage[version=4]{mhchem}
\usepackage{mathtools,xcolor}
\usepackage[normalem]{ulem}
\usepackage{braket}
\usepackage{siunitx}

\usepackage[T1]{fontenc}

\DeclareMathOperator{\Tr}{Tr}
\DeclareMathOperator{\tr}{tr}
\let\Re\relax

\DeclareMathOperator{\Re}{Re}

\newcommand{\ov}{\overline}

\renewcommand{\v}[1]{\bm{#1}}

\DeclarePairedDelimiter\abs{\lvert}{\rvert}%
\DeclarePairedDelimiter\norm{\lVert}{\rVert}%

\def\br{{\bm{r}}}
\def\bR{{\bm{R}}}
\def\bk{{\bm{k}}}
\def\bG{{\bm{G}}}
\def\bnabla{{\bm{\nabla}}}
\def\A{{\mathcal{A}}}
\def\B{{\mathcal{B}}}
\def\bz{{\mathrm{BZ}}}

\newtheorem{lemma}{Lemma}

\newcommand{\PRLsec}[1]{\textbf{\emph{#1---}}}

\begin{document}
\title{Higher Vortexability: Zero Field Realization of Higher Landau Levels}
\author{Manato Fujimoto}
\affiliation{Department of Physics, Harvard University, Cambridge, Massachusetts 02138, USA}
\affiliation{Department of Applied Physics, The University of Tokyo, Hongo, Tokyo, 113-8656, Japan}
\author{Daniel E. Parker}
\affiliation{Department of Physics, University of California at Berkeley, Berkeley, CA 94720, USA}
\affiliation{Department of Physics, University of California at San Diego, La Jolla, California 92093, USA}
\author{Junkai Dong}
\affiliation{Department of Physics, Harvard University, Cambridge, Massachusetts 02138, USA}
\author{Eslam Khalaf}
\affiliation{Department of Physics, Harvard University, Cambridge, Massachusetts 02138, USA}
\author{Ashvin Vishwanath}
\affiliation{Department of Physics, Harvard University, Cambridge, Massachusetts 02138, USA}
\author{Patrick Ledwith}
\affiliation{Department of Physics, Harvard University, Cambridge, Massachusetts 02138, USA}
\date{\today}

\begin{abstract}

The rise of moir\'{e} materials has led to experimental realizations of integer and FCIs in small or vanishing magnetic fields. At the same time, a set of minimal conditions sufficient to guarantee an Abelian fractional state in a flat band were identified, namely ``ideal" or ``vortexable" quantum geometry. Such vortexable bands share essential features with the LLL, while excluding the need for more fine-tuned aspects such as flat Berry curvature. A natural and important generalization is to ask if such conditions can be extended to capture the quantum geometry of higher Landau levels, particularly the first (1LL), where non-Abelian states at $\nu = 1/2,2/5$ are known to be competitive. The possibility of realizing these states at zero magnetic field, and perhaps even more exotic ones, could become a reality if we could identify the essential structure of the 1LL in Chern bands. In this work, we introduce a precise definition of 1LL quantum geometry, along with a figure of merit that measures how closely a given band approaches the 1LL. Periodically strained Bernal graphene is shown to realize such a 1LL structure even in zero magnetic field.
\end{abstract}

\maketitle

Fractional Chern Insulators (FCIs)~\cite{kapit2010exact, parameswaran2013fractional,bergholtz2013review,Neupertreview_2015,Liu_review_2023,NeupertFQHZeroField,sheng2011fractional,regnault2011fractional,tang2011high,sun2011nearly,Roy2014} realize the fractionalized charges and anyonic statistics associated with fractional quantum Hall physics in Chern bands. Such states do not require a magnetic field and can potentially form at much higher temperatures~\cite{spanton2018observation,xie2021fractional,cai2023signatures,zeng2023thermodynamic,park2023observation,xu2023observation,lu2023fractional}. They were observed in Harper-Hofstadter bands~\cite{spanton2018observation} at high magnetic field, and in the partially filled Chern bands of twisted bilayer graphene with the help of a small magnetic field~\cite{xie2021fractional}. Not only does the field reduce the bandwidth~\cite{parker2021field}, but it was also argued to improve the wavefunctions of the Chern band, pushing them to an ``ideal" or ``vortexable" limit where the band becomes a modulated version of the lowest LL (LLL)~\cite{ledwith2020fractional,ledwith2021strong,ozawa2021relations,mera2021kahler,wang2021chiral,wang2021exact,ledwith2022family,wang2022hierarchy,Dong2023Many,wang2023origin,ledwith2023vortexability}. More recently, Abelian FCIs at zero magnetic field, dubbed fractional quantum anomalous Hall (FQAH) states, were observed in twisted \ce{MoTe2}~\cite{cai2023signatures,zeng2023thermodynamic,park2023observation,xu2023observation}, following theoretical works showcasing small bandwidths and numerical works finding FCI ground states~\cite{wu2019topological,li2021spontaneous,crepel2023anomalous,dong2023composite,PhysRevLett.131.136501,hu2023hyperdeterminants,song2023intertwined,mao2023lattice,yu2023fractional,morales2023magic,jia2023moir,xu2024maximally,crepel2023chiral,wang2024fractional,li2024contrasting}, and also in pentalayer graphene~\cite{lu2023fractional,dong2023theory,zhou2023fractional,dong2023anomalous,herzog2023moir,kwan2023moir,guo2023theory}. As these zero field devices improve, they will likely run the gamut of Abelian topological orders found in the LLL.

\begin{figure}
\begin{center}
\includegraphics[width=0.25\textwidth]{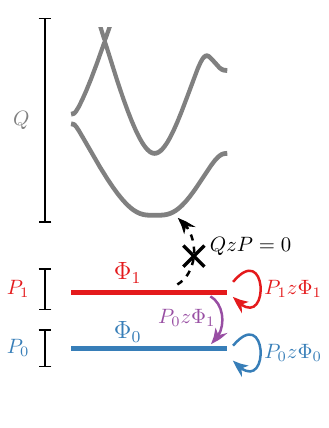}
\caption{Illustration depicting the definition \eqref{eq:defn_1vortexable}, of a first vortexable band, with band projector $P_1$.}
\label{fig_TBG_band_QG}
\end{center}
\end{figure}

To date, experimental realizations of non-Abelian topological order are the almost-exclusive province of the first LL. Here, we define the lowest Landau level as the zeroth LL and the subsequent level as the first LL.
Namely, the half-filled first LL in \ce{GaAs}~\cite{willett1987observation,pan2011impact,eisenstein2002insulating} and Bernal graphene~\cite{ki2014observation,kim2019even,zibrov2017tunable,li2017even,huang2022valley} realize a Pfaffian-type state with non-Abelian particle statistics\cite{moore1991nonabelions}. 
Also, the plateau at $\nu=2+2/5$ in \ce{GaAs}~\cite{Kumar2010} may be a non-Abelian phase harboring Fibonacci anyons~\cite{read1999beyond}, a scenario that has received numerical support ~\cite{read1999beyond,Zaletel}. 
Beyond their immense theoretical importance, such states have practical utility as a platform for fault-tolerant quantum computation~\cite{nayak2008non,mong2014universal}. Realizing non-Abelian states in Chern bands, at energy scales unachievable with external magnetic field, would be especially useful in this context.
This raises the question --- are there Chern bands at zero magnetic field analogous to the {\em first} LL?

In this work we answer this question by providing a precise definition of what it means for a band to be similar to the first LL. 
Our definition makes use of ``vortexability" applied to multi-band systems. A Chern band has first LL structure if we can find a partner band with a zeroth LL structure such that the two bands taken together are vortexable.
Vortex attachment applied to both bands produces {\em exact} Jain-like~\cite{jaincf} ground states in the limit of short-range repulsive interactions, without the need for further projection when the bands are degenerate. In general, the partner band is not required to be nearby in energy. We argue that single-band definitions, such as those based on the Fubini-Study metric used for the LLL case~\cite{roy2014band}, cannot capture first LL structure. Below we motivate and explicate our definition, then provide a concrete example in zero-field superlattice system, periodically strained Bernal graphene.
Our construction is readily extendable to higher Chern bands, by forming $(n+1)$ Chern band complexes to capture the geometry of the `$n$th' LL.


\PRLsec{LLL-type Quantum Geometry Review} 

We begin by reviewing the point of view that has recently emerged on ``LLL band geometry," through \textit{ideal}, or \textit{vortexable}, Chern bands~\cite{ledwith2020fractional,wang2021exact,ledwith2021strong,ozawa2021relations,mera2021kahler,mera2021engineering,varjas2022,ledwith2022family,wang2022hierarchy,ledwith2023vortexability,mera2023uniqueness,estienne2023ideal,Dong2023Many,wang2023origin,sarkar2023symmetry, liu2024broken}, which satisfy the following equivalent criteria. (i) the vortexability condition: $z \psi = P z \psi$, where $P$ is the projector onto the band(s) of interest, $\psi = P \psi$ is a wavefunction in those bands, and $z = x+iy$ is the vortex function \cite{ledwith2022family,ledwith2023vortexability} (ii) the trace condition: $\tr g(\bk) = \Omega(\bk)$~\cite{Parameswaran_2013,Roy2014, jackson2015geometric, bauer2016quantum}, where $g_{\mu \nu}(\bk)$ is the Fubini-Study metric and $\Omega(\bk)$ is the Berry curvature.  (iii) momentum-space holomorphicity: the cell-periodic states $u_{\bk a}(\br) = e^{-i\bk \cdot \br} \psi_{\bk a}(\br)$ can be chosen to be holomorphic in $k=k_x+ik_y$ (here ``$a$" is a band index)~\cite{claassen2015position,tarnopolskyorigin2019,mera2021kahler}. The vortexability condition (i) was recently identified as the crucial ingredient required to generalize Trugman-Kivelson pseudopotential argument to Chern bands which allows obtaining exact ground states for short-range interactions. Vortexable bands can be defined with more general vortex functions $z(\br) \neq x+iy$ \cite{ledwith2023vortexability} (equivalent to generalized ideal bands with $g, \Omega$ from modified unit cell embeddings \cite{SimonRudner,liu2024theory}) for which all our conclusions still hold. In what follows, we will assume the standard vortex function $z=x+iy$ and use vortexability and ideality interchangeably.(see Appendix~\ref{sec_app_multiband_QG} in detail)

 For zero dispersion vortexable bands, Laughlin-like trial states constructed by attaching vortices (i.e. Jastrow factors) $\prod_{i<j} (z_i - z_j)^{2m}$ to integer filling states are exact many-body ground states in the limit of short range repulsive interactions (Trugman-Kivelson type)~\cite{TrugmanKivelson1985,pokrovskySimpleModelFractional1985,ledwith2020fractional,ledwith2023vortexability}. Thus, these generalizations of the LLL provide a frustration-free route to realizing FCIs. Higher LLs are not vortexable and have distinct interacting phases; note that mapping interactions $V_{\bm q}$ in the $n$'th LL to a short range LLL interaction $\tilde{V}_{\bm q}$ requires $V_{\bm q} = \tilde{V}_{\bm q}/|L_n(\abs{\bm q}^2 \ell_B^2/2)|^2$, where $\ell_B$ is the magnetic length and $L_n$ is the $n$'th Laguerre polynomial\cite{foglerStripeBubblePhases2001a}. In contrast to naive intuition and early work on FCIs, homogeneous Berry curvature is not a requirement.

Wavefunctions for $C=1$ ideal bands have the form~\cite{ledwith2020fractional,wang2021exact}
\begin{equation}
\label{eq:ideal_wavefunction}
   \psi(\br) = f(z) \mathcal{N}_0(\br); \quad \mathcal{N}_{0}^{l}(\br) = e^{-K(\br)} \zeta^l(\br)
\end{equation}
where $f(z)$ is holomorphic, $\zeta^l(\br)$ is an orbital-space spinor (with layer index $l$, e.g.), and $\phi(\br) = f(z) e^{-K(\br)}$ is a zero mode of a Dirac particle in the inhomogeneous magnetic field $B(\br) = \nabla^2 \Re K(\br)$. Due to its normalization $\sum_l \abs{\zeta^l(\br)}^2 = 1$ the spinor part does not influence the energy of the band under density-density interactions, and can thus be largely ignored for our purposes.
Here $\phi$ and $\zeta$ are symmetric under magnetic translations with opposite phase. The class of $C=1$ vortexable bands therefore has the same analytic structure as the LLL in an inhomogenous magnetic field and metric that breaks continuous translation symmetry~\cite{ledwith2020fractional,estienne2023ideal}. Vortexability is thus a sufficient condition for a $C=1$ band to be ``analogous to the LLL".

\PRLsec{First Landau Levels Are Diverse} To extend the success of ideal bands to the first LL, one could imagine imposing band geometric criteria to mimic a 1LL features. 
For example, the first LL of a $p^2/2m$ particle has $\tr g = 3 \Omega$~\cite{ozawa2021relations}. However, this and other approaches fail to capture the wide spectrum of ``first LLs" in multilayer graphene. To wit, we recall Bernal graphene at $B > 0$: 
\begin{equation}
\label{eq_BG_LL_Ham}
H_{\rm BG}=\left(\begin{array}{cc}0 & D_{1}^{\dagger} \\ D_{1} & 0\end{array}\right), \quad D_{1}=\left(\begin{array}{cc}D_{0} & \gamma_{1} \\ 0 & D_{0}\end{array}\right),
\end{equation}
written in the chiral basis $(|A,1\rangle,|A,2\rangle,|B,1\rangle,|B,2\rangle)$. Here $D_{0}=2v\left(-i \bar{\partial}_{z}-\bar{A}_{z}\right)=\sqrt{2} v l_B^{-1} \hat{a}$ is proportional to the annihilation operator and $\gamma_1$ is the interlayer coupling. There are two bands of zero modes; $\ker D_1$ is spanned by
\begin{equation}
\label{eq_BG_LL_States}
\Phi_{0}=\left(\begin{array}{c}\psi_{0} \\ 0\end{array}\right), \quad \Phi_{1}=\left(\begin{array}{c}v^{-1} \ell_B \gamma_1 \hat{a}^{\dagger} \psi_{0} \\ -\sqrt{2}  \psi_{0}\end{array}\right),
\end{equation}
forming a zeroth and first LL respectively, where $\psi_{0}(\br) = f(z) e^{-\abs{z}^2/4 \ell_B^2}$ in symmetric gauge with $f(z)$ holomorphic.

Direct computation shows that, as $\gamma_1$ varies from $0$ to $\infty$, the ratio $\tr g_{1}/\Omega_{1}$ of band $\Phi_{1}$ varies continuously from $1$ to $3$. However, the wavefunctions have a similar analytic structure throughout. Even if one insists the first LL corresponds only to the $\gamma_1 \to \infty$ limit, with $\tr g_{1} = 3 \Omega_{1}$, the condition $\tr g = 3 \Omega$ is also satisfied by the second LL of trilayer graphene for a particular value of $v^{-1}\ell_B\gamma_1$. The single-band Berry curvature and quantum metric therefore appear insufficient to specify a first LL structure.

The fact that the first $n$ LLs are ideal, when taken together, was first noticed in \cite{ozawa2021relations} by explicitly calculating $\tr g = \Omega$ for a $p^2/2m$ particle. This property that adding vortices to the first LL gives components in both first and zeroth LLs is the crux of our generalization.

\PRLsec{Definition of First LL Structure} 
We now precisely define the structure bands must satisfy to be ``first LL analogues". 
A Chern band, with projector $P_1$, is \textit{first vortexable} if the following two conditions are satisfied:
\begin{enumerate}
\item ($\exists$ Partner Band) There exists an orthogonal vortexable band with projector $P_0$ 
(i.e. $P_0 z \Phi_0~=~z \Phi_0$ for all $P_0 \Phi_0 = \Phi_0$) such that
\begin{align}
\label{eq:defn_1vortexable}
(P_1 + P_0) z\Phi_1 &= z \Phi_1
\end{align}
for all states $P_1\Phi_1 = \Phi_1$.
\item (Indecomposability) there is no alternative basis of wavefunctions $(\Phi_{0, \bk}', \Phi_{1, \bk}')$, where each band is vortexable, that also spans the two band subspace of $P_0 + P_1$.
\end{enumerate}
The additional band, described by the projector $P_0$, functions as an effective zeroth LL. Intuitively, acting with $z$ either keeps the wavefunction within the `first' band $P_1$ or lowers it to the `zeroth' band $P_0$ --- but does not generate components in any other bands (see Fig. \ref{fig_TBG_band_QG}a). The second condition is in principle necessary to avoid subtle situations where the first condition can nominally be realized, but the system should be thought of as a collection of two separate, individually vortexable, bands instead. Given a single Bloch band $\Phi_1$, the partner band is $\Phi_0 \propto (1-P_1)z \Phi_1$ (see Appendix~\ref{sec_app_partner} for algorithmic construction), so Eq.~\eqref{eq:defn_1vortexable} can always be explicitly checked.

A $C=1$ band is first-vortexable if, and only if, its wavefunctions have the form
\begin{equation}
    \Phi_{1}(\br) = Q_0\left[ \left(\overline{z}f(z) - 2 B_0^{-1} f'(z) \right) \mathcal{N}_{0}(\br) + f(z) \mathcal{N}_{1}(\br)  \right].
    \label{eq:first_vortex_genform}
\end{equation}
Here $Q_0 = 1-P_0$, $\mathcal{N}_{0,1}$ are spinor-valued [\textit{cf}. Eq.~\eqref{eq:ideal_wavefunction}], and $P_0$ is the projector onto the ``zeroth-LL" wavefunctions $f(z)\mathcal{N}_0(\br)$. $B_0=\frac{2 \pi}{A_{\mathrm{UC}}}=\frac{A_{\mathrm{BZ}}}{2 \pi}$, and $A_{\mathrm{UC}}$ and $A_{\mathrm{BZ}}$ are the areas of the unit cell and the Brillouin Zone (BZ), respectively. 
Appendix~\ref{sec_app_first_order_vortexability_tiral} proves that Eq. \eqref{eq:first_vortex_genform} implies first-vortexability; the reverse implication will be proven in \cite{classifyInPrep}. In Appendix~\ref{sec_app_analytical_example}, we derive a zero-field concrete form of Eq.\eqref{eq:first_vortex_genform} in an analytically solvable model motivated by magic angle twisted bilayer graphene.

First vortexability has immediate many-body implications. Consider a first vortexable band that is flat and degenerate with its vortexable counterpart, as in Bernal graphene and in a upcoming example. Let $\Psi_{\nu=2}$ be the many-body state where both bands are full. Now consider the vortex attached state $\Psi_v = \prod_{i<j} (z_i - z_j)^{2s} \Psi_{\nu=p}$ which has filling $p/(2ps+1)$, where $p=2$ (note, general $p$ follows analogously for $(p-1)$'th vortexability). Thanks to vortexability of the combined bands, attaching this vortex factor keeps the final state within the combined bands. The above construction is similar to that of LLL Jain states~\cite{jaincf}, where projection into the LLL is strictly speaking required at the end. Here however, \textit{LLL projection is unnecessary}; $\Psi_v$ is an exact zero mode of short-range interactions~\cite{TrugmanKivelson1985,ledwith2020fractional,ledwith2023vortexability}.

We also introduce a refinement of our criterion, which is specifically geared towards realizing states unique to the first LL. While we motivated our definition through the inclusion of Bernal graphene for any value of the interlayer tunneling, as a byproduct we have included bands that for all practical purposes function as LLLs; Consider the decoupled $\gamma_1 \to 0$ limit, which adiabatically connects Jain-like states with layer-singlet Halperin-like states at $\gamma_1=0$. At half filling of the first-vortexable band, however, $\gamma_1$ must be sufficiently large for a Pfaffian to outcompete the composite Fermi liquid at half filling \cite{ZhuWidelyTunable2020}. It turns out that the $\gamma_1 \to \infty$ limit, or more generally $\mathcal{N}_{1} \to 0$, has some additional structure in non-Abelian band geometry that leads to a natural quantification of how close a band is to the ``maximal" first LL. In Appendix~\ref{sec_app_max_QGT}, we show that for $\mathcal{N}_{1} \to 0$, the non-Abelian Berry curvature (in the space of the zeroth and first LLs) becomes rank deficient, with only one nonzero entry $\hat{\Omega}_{11}$. We  therefore define a ``maximality index"
\begin{equation}
\label{eq_Mbk}
    M(\bk) = \frac{\abs{\lambda_1(\bk) - \lambda_2(\bk)}}{\lambda_1(\bk) + \lambda_2(\bk)},
\end{equation}
where $\lambda_{1,2}$ are the eigenvalues of the $2 \times 2$ non-Abelian Berry curvature. As $\gamma_1$ varies from $0\to \infty$ in Bernal graphene, $M(\bk) = \gamma_1^2/(2 (v \ell_B^{-1})^2+\gamma_1^2)$ increases from $0 \to 1$ (see Appendix~\ref{sec_app_Mk}). 
First vortexable bands with $M(\bk) \to 1$ (i.e. $\mathcal{N}_1 \to 0$) are effectively 1LLs with periodic modulation $\norm{\mathcal{N}_{0}(\br)}/e^{-B_0\abs{\br}^2/4}$. We expect them to host Pfaffian-type topological orders.


\begin{figure}
\begin{center}
    \includegraphics[width=1.0 \hsize]{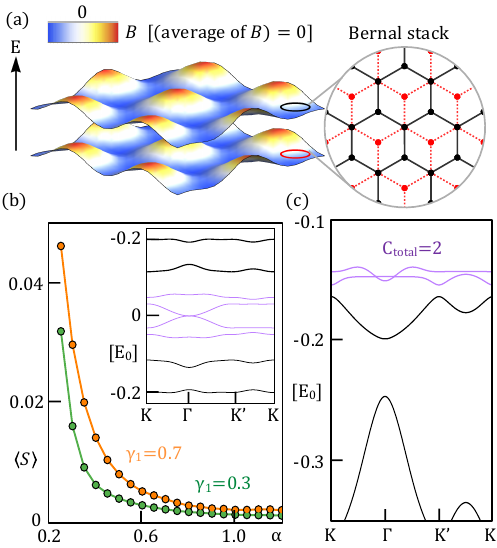}
    \caption{(a) Bernal graphene with vertical buckling and a strain field, giving a periodic psuedomagnetic field $B$ (color map) with zero average. (b) Average overlap deviation between numerical and analytical approximate zero modes with $\eta=0.5$, $\langle S \rangle_{\rm BZ}$, versus $\alpha$ at $\beta=0$. Inset: bandstructure for $(\alpha,\beta,\gamma_1)=(0.5,0,0.5)$ with low-energy modes $E_{0,1}$ in purple. (c) Bandstructure at  $(\alpha,\beta,\gamma_1)=(0.4,0.068,1)$. Displacement field $\beta > 0$ isolates the two low-energy Chern bands (purple).}
    \label{fig_SBG_model}
\end{center}
\end{figure}

\PRLsec{Periodically Strained Bernal Graphene}
 We will now show how a first vortexable band appears amongst the low energy bands of periodically strained bilayer graphene~\cite{wan2023nearly}. We will take a similar approach to Ref. \cite{gao2023untwisting} which studied periodically strained monolayer graphene with a $C_2$-breaking substrate that induces $C_2$ breaking pseudomagnetic field 
\begin{equation}
\mathcal{B}(z, \bar{z})=\mathcal{B}_{0} \sum_{l=0}^{5} e^{i \bm{G}_{l} \cdot \bm{r}}=\mathcal{B}_{0} \sum_{l=0}^{5} e^{\frac{i}{2}\left(G_{l} \bar{z}+\bar{G}_{l} z\right)}
\end{equation}
where $\bm{G}_{l}=R_{2\pi l / 6} \bm{G}_{0}, \bm{G}_{0}=\frac{4 \pi}{\sqrt{3} L_{M}}(1,0)$ are smallest reciprocal lattice vectors in terms of the period $L_M$; the addition of higher harmonics is not expected to change our conclusions~\cite{gao2023untwisting}. We note that $\B(\br) = \B(-\br)$, within a valley, implies that the pseudomagnetic field is $C_2$-odd. Furthermore, $\B(r)$ averages to zero such that the bands we will obtain are Bloch bands, not LLs. 
The periodic strain can be realized by placing monolayer graphene on a lattice of nanorods~\cite{jiang2017visualizing}, or by spontaneous buckling of a graphene sheet on $C_2$ breaking substrates such as \ce{NbSe2}~\cite{maoEvidenceFlatBands2020}. We will also include a spatially-varying scalar potential with the same periodicity proportional to $V(\br) = \sum_l e^{i \v G_l \cdot \br}$ and of strength $V_0$. The scalar potential could come from patterned gates~\cite{forsythe2018band,shi2019gate,cano2021moire,guerci2022designer,ghorashi2023multilayer}, or an electric field that couples to the buckling height.

The (de-dimensionalized) Hamiltonian for AB-stacked graphene's $K$ valley is then
\begin{equation}
\label{eq_SBG_Ham}
\begin{aligned}
H_{\rm SBG} &=
E_0\left(\begin{array}{cc}
 h   & \gamma_1 \sigma_{+} \\
 \gamma_1 \sigma_{-}  &  h
\end{array}\right),\\
h & = \v \sigma \cdot [-i \v \nabla + \alpha \v \A] - \beta V(\br),
\end{aligned}
\end{equation}
where $E_0 = \hbar v_F \abs{\v G_0}$, $\sigma_\pm = \frac{1}{2}(\sigma_+ \pm i \sigma_-)$, $\alpha = 1/\ell_B^2\abs{G_0}^2$, $\beta=V_0/E_0$, $\B_0 = \hbar/e\ell_B^2$, $\A = \mathcal{A}_x + i A_y = \sum_l e^{2\pi i/l} e^{i \v G_l \cdot \v r}$~We will neglect the interlayer potential for simplicity, consistent with patterned dielectric gates sandwiching the sample..
We take interlayer tunnelling $E_0 \gamma_1 \approx 370$ meV and, following ~\cite{maoEvidenceFlatBands2020}, $E_0 \approx \SI{300}{meV}$. The bandstructure, Fig.~\ref{fig_SBG_model}(b) inset, has a low energy subspace of four bands (purple) with total bandwidth $O(e^{-6\alpha})$. In the chiral limit $\beta=0$, $H_{\rm SBG}/E_0$ takes the form of Eq.~\eqref{eq_BG_LL_Ham} but now with $D_0 = -2i \ov{\partial} + \alpha \A(\br)$.

We explicitly construct trial wavefunctions on the $A$ sublattice for a zeroth and first vortexable band that --- up to exponentially small corrections $O(e^{-6\alpha})$ --- are zero modes of $D_1$. The two trial wavefunctions are Chern bands whose layer spinors take the form
\begin{equation}
\label{eq_strained_bernal_approxzero_main}
\Phi_{0,\bm{k}}(\bm{r}) =\left(\begin{array}{c}\psi_{0,\bm{k}}(\bm{r}) \\ 0\end{array}\right),
\Phi_{1,\bm{k}}(\bm{r})=\left(\begin{array}{c}  \psi_{1,\bm{k}}(\bm{r}) \\ 2i \gamma_1^{-1}\psi_{0,\bm{k}}(\bm{r})\end{array}\right),
\end{equation}
with precise expressions given in Appendix~\ref{sec_app_trial_wavefunction_strainedgraphene}. 
The construction of \eqref{eq_strained_bernal_approxzero_main} follows the same reasoning as Ref. \cite{gao2023untwisting}, making use of $[D_0,z]=0$ and $\psi_\Gamma(\br = 0) = e^{-6 \alpha} \approx 0$, where $V(\br = 0) = 6$.
In Appendix~\ref{sec_app_trial_wavefunction_strainedgraphene} we also construct two topologically trivial near zero modes of $D_1^\dagger$ for the $B$-sublattice. We now use these trial wavefunctions to identify a first vortexable band of $H_{\mathrm{SBG}}$ and isolate it from the rest of the spectrum.

\begin{figure}
\begin{center}
    \includegraphics[width=1.0 \hsize]{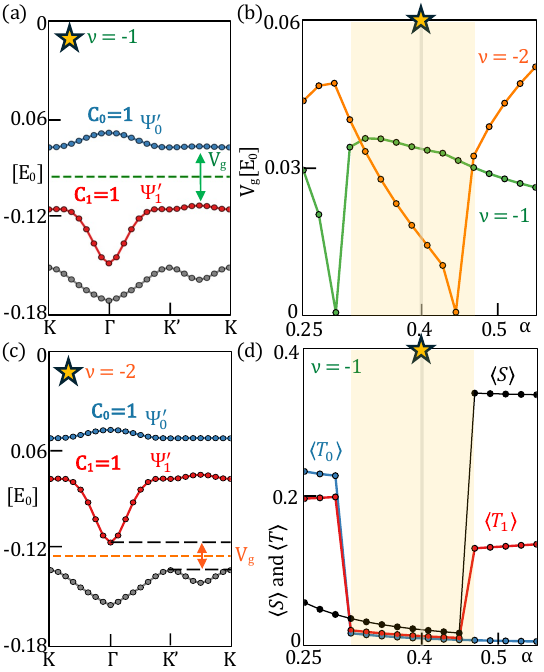}
    \caption{
    (a),(c): SCHF bandstructures for $\nu=-1$ and $\nu=-2$, respectively, with parameters corresponding to Fig.~\ref{fig_SBG_model}(c), and $(d,\epsilon_r)=(\SI{250}{\angstrom},15)$. The dashed lines represent the Fermi energy. (b) The band gaps $V_g$ for $\nu=-1$ and $\nu=-2$ are plotted with green and orange lines, respectively. 
    (d) Average overlap deviation between numerical and analytical eigenstates with $\eta=0.5$ at $\nu=-1$ as a function of $\alpha$. Here  $\langle T_{0} \rangle_{\rm BZ}$ (blue), and $\langle T_{1} \rangle_{\rm BZ}$ (red) correspond to the zeroth and first vortexable bands, whereas $\langle S \rangle_{\rm BZ}$ is tow-band complex.
    }
    \label{fig_SBG_band_QG}
\end{center}
\end{figure}

We now numerically verify that the low-energy subspace of $H_{\mathrm{SBG}}$ on the $A$-sublattice is spanned by the trial wavefunctions Eq.\eqref{eq_strained_bernal_approxzero_main} up to $O(e^{-6\alpha})$ corrections. As the energy eigenstates are mixed between the sublattices (at $\beta = 0$), they cannot correspond to the trial wavefunctions directly. Instead, we use the sublattice basis $\Psi^{A,B}_{n,\bk} = \frac{1}{\sqrt{2}}( \Psi_{E_n,\bk} \pm \sigma_z \Psi_{E_n, \bk})$, where $n=0,1$ labels the two eigenstates of smallest $E_n > 0$, and $\sigma_z \Psi_{E_n} \propto \Psi_{-E_n}$. The overlap deviation between the span of the trial wavefunctions $P_{\Phi \bk}=\sum_n |\Phi_{n,\bk}\rangle \langle \Phi_{n,\bk}|$ and the $A$-subspace $P_{\Psi \bk}=\sum_n |\Psi^{A}_{n, \bk}\rangle \langle \Psi^{A}_{n, \bk}|$ is quantified by $ S_\bk[\Phi_{\bk},\Psi^A_{\bk}] = 1-\mathrm{Tr}[P_{\Phi \bk}P_{\Psi \bk}]/2$. The trial wavefunctions span the $A$-subspace if and only if the Brillouin zone average vanishes, $\braket{S}_{\mathrm{BZ}} \approx 0$. Fig.~\ref{fig_SBG_model}(b) shows $\langle S \rangle_{\mathrm{BZ}}$ decreases as $e^{-6\alpha}$, giving overlaps $>0.99$ whenever $\alpha >0.2$. We therefore conclude that the low-energy $A$-sublattice subspace is spanned by ``almost" vortexable and first vortexable bands, satisfying the definition up to small $O(e^{-6\alpha})$ corrections.

Finally, we show how interactions can spectrally isolate the almost first-vortexable band. Chiral symmetry is lifted by displacement fields $\beta > 0$, breaking the Dirac cone and separating out a complex of two bands $\Psi^A_{(0,1)}$ with $C_{\rm total}=2$. The complex's total bandwidth is minimized at $\beta=0.068$ [Fig \ref{fig_SBG_model}(c)], but the two bands within it are not isolated without adding interactions. We now add gate-screened Coulomb interactions in the standard way, and study the resulting self-consistent Hartree-Fock (SCHF) bandstructures at integer fillings (details in Appendix~\ref{sec_app_SCHF}). We find interactions split the complex into two isolated bands $\Psi_0'$ and $\Psi_1'$ that approximately are vortexable and first-vortexable, respectively. The overlap deviation $\braket{S}$, shown in Fig. \ref{fig_SBG_band_QG}(d), decreases as $e^{-5.6 \alpha}$. This is directly analogous to Bernal graphene in field, where interactions split the degenerate zeroth and first LLs since the more-concentrated LLL wavefunctions have less exchange energy~\cite{MacdonaldOrig,MacdonaldReview,MacdonaldRecent}.

Fig.~\ref{fig_SBG_band_QG} shows SCHF bandstructures at fillings $\nu=-1$ and $-2$, each of which shows two isolated $C=1$ bands. At $\nu=-2$, the SCHF bandgap closes near $\alpha \approx 0.45$, enabling a topological transition $C=(1,1) \to (1,0)$ by mixing with the third band. Conversely, the top two bands hybridize to give isolated bands with $C=(2,0)$ below $\alpha \approx 0.3$. Hence the two $C=1$ bands are isolated over a range of $\alpha$ for both $\nu=-1$ \& $-2$.

To confirm vortexability, we compare the numerical wavefunctions to the analytical ans\"atze $\Phi_{(0,1)}$ from Eq.~\eqref{eq_strained_bernal_approxzero_main}. Fig.~\ref{fig_SBG_band_QG}(d) shows $\langle T_{n}\rangle$, the average of $1-\left|\braket{\Psi_{n\bk}^{A} | \Phi_{n\bk}}\right|$ over the BZ and the two band overlap deviation $\langle S \rangle$ defined above. In the regime $0.3 < \alpha < 0.45$ with $C=(1,1)$, the overlap deviation decreases as $O(e^{-5.6\alpha})$, giving deviations $<1\%$ for $\alpha > 0.38$. The numerical wavefunctions therefore match the analytic ones to extremely good precision, and the latter are almost vortexable and first vortexable. We compute the maximality index, Eq.~\eqref{eq_Mbk}, to be $\langle M \rangle_{\rm BZ} \approx 0.79$ at $\alpha=0.4$ and dimensionless tunneling $\gamma_1=1$. This is close to the optimal value of $1$.

In this work we have proposed ``first vortexability" --- a precise definition of when a Chern band has the essential character of the first LL. First vortexable bands are defined by their quantum geometric data, and each requires a partner band analogous to the zeroth LL. Our examples, both analytic and physically reasonable, suggest such bands are experimentally realizable. Future work will investigate the many-body physics of partially-filled first vortexable bands, particularly at half filling where non-Abelian states are likely.

\begin{acknowledgements}
We thank Jie Wang, Bruno Mera, Tomohiro Soejima, and Mike Zaletel for enlightening discussions. D.E.P. is supported by the Simons Collaboration on UltraQuantum Matter, which is a grant from the Simons Foundation. D.E.P. is supported by startup funds from the University of California, San Diego. M.F. is funded by JSPS KAKENHI Grant No. JP23KJ0339 and JP24K16987, an Energy Frontier Research Center at the Ames National Laboratory. Work at the Ames National Laboratory is supported by the U.S. Department of Energy (DOE), Basic Energy Sciences (BES) and is operated for the U.S. DOE by Iowa State University under Contract No. DE-AC02-07CH11358. A. V. and M. F. are funded by Center for the Advancement of Topological Semimetals CATs.

\end{acknowledgements}

\appendix


\section{Multi-band Vortexable Band Geometry}
\label{sec_app_multiband_QG}

In this section we review ideal, or vortexable, band geometry for multiple bands and study the non-Abelian quantum geometric tensor in this context for later use. Consider a isolated band or set of bands $\psi_{\bk a}(\br)$ and corresponding periodic parts $u_{\bk a}(\br) = e^{-i\bk \cdot \br} \psi_{\bk a}(\br)$. Their non-Abelian quantum geometric tensor (QGT) is given, in an orthonormal basis, by
\begin{equation}
\begin{aligned}
    \hat{\eta}^{\mu \nu}_{ab} &= \bra{\partial_{k_{\nu}} u_{\bk b} } Q(\bk) \ket{\partial_{k_{\mu}} u_{\bk a}},\\
    Q(\bk) & = 1 - \sum_{a} \ket{u_{\bk a}} \bra{u_{\bk a}}
    \end{aligned}
\end{equation}
where $a,b$ are band-indices. 

We will denote its trace by removing its hat, or with a capital $\Tr$ (and use $\tr = \sum_{\mu}$ for spatial indices), i.e. 
\begin{equation}
    \eta^{\mu \nu} = \Tr \hat{\eta}^{\mu \nu} = \sum_a \eta^{\mu \nu}_{aa}.
    \label{eq:QGTsupp}
\end{equation}
The non-Abelian Berry curvature and metric are defined as the symmetric and antisymmetric components in the spatial indices 
\begin{equation}
\begin{aligned}
\hat{g}^{\mu \nu}(\bk) & = \frac{1}{2}(\hat{\eta}^{\mu \nu} + \hat{\eta}^{\nu  \mu}), \\
    \hat{\Omega}_{ab} \varepsilon^{\mu \nu} & = i(\hat{\eta}^{\mu \nu} - \hat{\eta}^{\nu \mu}).
    \end{aligned}
\end{equation}

We now give five equivalent definitions of vortexability with vortex function $z = x+iy$, or ideal band geometry
\cite{ledwith2023vortexability,mera2021kahler,ozawa2021relations,wang2021exact,ledwith2022family}.
\begin{enumerate}
\item[(V1)] 
\label{V1} The bands are vortexable with vortex function $z = x+iy$. That is, $z \psi = P z \psi$, where $\psi = P \psi$ is any wavefunction in the band and $P$ is the band projector.

\item[(V2)] 
	\label{V2}The inequality $\tr g \geq \Omega$, or $\tr \Tr \hat{g} \geq \Tr \hat{\Omega}$, is saturated.
\item[(V3)] 
\label{V3}
There is a gauge choice such that $\ket{\tilde{u}_{k a}} = S_{\bk} \ket{u_{\bk a}}$ is a holomorphic function of $k = k_x + i k_y$. Here $S_{\bk}$ is an invertible matrix that is generically non-unitary, such that the holomorphic gauge wavefunctions are generically not normalized and non-orthogonal.

\item[(V4)] 
\label{V4}The quantum metric factorizes as 
$\hat{\eta}^{\mu \nu}_{ab} = \frac{1}{2} \hat{\Omega}_{ab}(\delta^{\mu \nu} - i \varepsilon^{\mu \nu})$, where $\hat{\Omega}_{ab} = \bra{\partial_{k} u_{\bk b} } Q(\bk) \ket{\partial_{k} u_{\bk a}}$ is the non-Abelian Berry curvature and $2\partial_k = \partial_{k_x} + i \partial_{k_y}$ is the derivative with respect to $k = k_x + i k_y$
\end{enumerate}

We emphasize that the above equivalences only hold for vortexability with vortex function $z = x+iy$. Bands can be vortexable under vortex functions $z(\br)$ that are not equivalent to $x+iy$, including nonlinear functions, with similar implications for fractional Chern insulators under short range repulsive interactions. However, under mild physical constraints on the vortex function \cite{ledwith2023vortexability} it is always possible to find a change of coordinates $\phi(x,y) = (x',y')$ such that $z(\br) = x' + i y'$. Within these new coordinates, analogues of conditions (V2-4) may be derived that are equivalent to vortexability under $z(\br)$. Thus, while the two vortexable functions are \emph{physically} distinct, (for example, the Coulomb interaction is not invariant under the change of coordinates $\phi$) at a formal level it suffices to study vortexability under $x+iy$.

Ref. \cite{ledwith2023vortexability} showed the equivalence between (V1) and (V2), and reviewed the equivalence between (V2) and (V3) through an explicit construction of the required gauge transformation, which was previously shown to exist by Ref. \cite{mera2021kahler}. We note that the equivalence is specific to the vortex function $x+iy$; vortexability with more general vortex functions are equivalent to appropriately generalized versions of the trace condtition \cite{ledwith2023vortexability}.
It remains to show (V4) is equivalent to (V1-3). The implication $\rm{(V4)} \implies \rm{(V2)}$ follows immediately by computing $\tr g$ and $\Omega$ from the traced symmetric and antisymmetric parts of $\hat{\eta}^{\mu \nu}_{ab} = \frac{1}{2} \hat{\Omega}_{ab}(\delta^{\mu \nu} - i \varepsilon^{\mu \nu})$, respectively. We conclude by showing $\rm{(V3)} \implies \rm{(V4)}$. To see this, we use $(V3)$ to write $\ket{u_{\bk} }  = S_\bk \ket{\tilde{u}_{k}}$, where $S_\bk$ is an invertible matrix that is not necessarily unitary, and $\ket{\tilde{u}_{k}}$ is holomorphic. Then we have 
\begin{equation}
\begin{aligned}
    Q(\bk) \partial_{k_x} \ket{ u_{\bk a}} & =Q(\bk) (\partial_k + \ov{\partial}_k) S_\bk \ket{\tilde{u}_{k }} \\
    & = S_\bk Q(\bk) \partial_k\ket{\tilde{u}_{k a}}\\
    & = Q(\bk)  \partial_k\ket{u_{\bk a}}
    \end{aligned}
\end{equation}
where we used that the term coming from differentiating $S(\bk)$ is annihilated by $Q(\bk)$, so that $S_\bk$ commutes through $Q \partial_{k_\mu}$. Similarly we can write $Q(\bk) \partial_{k_y} \ket{ u_{\bk }} = i Q(\bk) \partial_{k} \ket{ u_{\bk }}$. Substituting into \eqref{eq:QGTsupp} then yields (V4).

\section{First vortexability of 1LL-type wavefunctions}
\label{sec_app_first_order_vortexability_tiral}

In the main text we claimed that a band is $C=1$ first-vortexable if, and only if, it has the form
\begin{equation}
    \Phi_{1}(\br) = Q_0\left[ \left(\overline{z}f(z) - 2 B_0^{-1} f'(z) \right) \mathcal{N}_{0}(\br) + f(z) \mathcal{N}_{1}(\br)  \right].
\end{equation}
This section will explicitly demonstrate that Eq.~\eqref{eq:first_vortex_genform} implies first-vortexability; the more-technical converse implication will be proven in later work \cite{classifyInPrep}. Immediately below we define the notation associated with Eq.~\eqref{eq:first_vortex_genform} and provide some intuition for its generality \cite{classifyInPrep}.
Afterwards we provide the proof that wavefunctions of the form \eqref{eq:first_vortex_genform} are first-vortexable.

Our framework for this section is wavefunctions $\Phi^\ell(\br)$ on the plane with an internal (``layer") index $\ell$. We assume there is a direct lattice\footnote{In the Landau level case, this is the ghost lattice of magnetic translations.} generated by $\bR_{1,2}$ and define an effective average magnetic field $B_0$ that would give one ``flux quantum" per unit cell: $B_0 A_{\text{uc}} = 2\pi$. We can now unpack Eq.~\eqref{eq:first_vortex_genform}. To span a whole band, the complex-valued functions $f(z)$ run over all holomorphic functions of $z=x+iy$. These generate the zeroth vortexable partner band
\begin{equation}
    \Phi_{0}(\br) = f(z) \mathcal{N}_{0}(\br),
    \label{eq:zero_vortex_genform}
\end{equation}
and we define a projector $Q_0$ onto wavefunctions orthogonal to Eq.~\eqref{eq:zero_vortex_genform}. The functions $\mathcal{N}_{0,1}^l(\br)$ are functions of the layer and real-space, but do not depend on the state in the band; the state is determined by the holomorphic function $f(z)$. 

The periodicity of $\mathcal{N}$ under translation by a lattice vector $\bR$ is given by
\begin{equation}
    \mathcal{N}_{0,1}(\br + \bR) = \eta_\bR e^{-\frac{i}{2} B_0 \bR \times (\br-\br_0) } e^{-\frac{B_0}{2} \bR \cdot (\br-\br_0) - \frac{B_0}{4}\abs{\bR}^2} \mathcal{N}_{0,1}(\br)
    \label{eq:translationofN}
\end{equation}
where $\eta_\bR = 1$ if $\bR/2$ is a lattice vector and $-1$ otherwise. The shift $\br_0$ represents a potential shift in the periodicity relative to the choice of origin; for simplicity we will choose our real-space origin so that $\br_0 \to 0$, which naturally occurs in all the models we consider. The factors $\mathcal{N}_{0,1}$ serve as analogs of the LLL Gaussian factor $e^{-B_0 \abs{z}^2}$ for normal translations as opposed to magnetic translations. Specifically, the two factors, $\eta_\bR e^{-\frac{i}{2} B_0 \bR \times \br }$, which may be regarded as the factor generated by the phase of $\mathcal{N}$, enable the rest of the wavefunction to be symmetric under magnetic translations with an average flux $B_0$ corresponding to one flux quantum per unit cell. The factor $e^{-\frac{B_0}{2} \bR \cdot \br - \frac{B_0}{4}\abs{\bR}^2}$, which may be regarded as the transformation of the magnitude of $\mathcal{N}$, is equivalent to the factor generated by the 
the LLL Gaussian $e^{-B_0 \abs{\br}^2/4}$ under $\br \to \br + \bR$. We note that we have chosen to interpret \eqref{eq:translationofN} through an analogy with symmetric gauge quantum Hall wavefunctions for convenience and ease of interpretation; the rule \eqref{eq:translationofN} may be interpreted differently in other gauges by combining the exponentials and re-splitting them in a different way. Finally we comment on translation-invariance. Since there is a lattice, we can always choose to write Eq.~\eqref{eq:first_vortex_genform} in a Bloch basis. In this case, the only $k$-dependence will be through $f(z) \to f_{k}(z)$. Below we give an explicit (but not unique) form of $f_k(z)$ in terms of (modified) Weierstrauss $\sigma$-functions, as these functions naturally appear in the symmetric-gauge lowest Landau level wavefunctions on the torus \cite{haldane2018modular}.

Let us briefly provide some intuition behind the generality of \eqref{eq:first_vortex_genform}, which will be proven in a follow up work \cite{classifyInPrep}. It may seem to the reader too restrictive that a quantum geometric condition like first vortexability can restrict the wavefunctions to the form \eqref{eq:first_vortex_genform}. However, \eqref{eq:first_vortex_genform} can be viewed as a generalization of \eqref{eq:zero_vortex_genform}, proved in \cite{wang2021exact,Dong2023Many}, which is equally restrictive. The restriction to wavefunctions of the form \eqref{eq:zero_vortex_genform},\eqref{eq:first_vortex_genform} can be viewed through the rigidity of holomorphic functions in $k_x + i k_y$. Indeed, condition (V3) implies that the corresponding periodic states $u_k = e^{-i \bk \cdot \br} \Phi_\bk(\br)$ can be written as holomorphic functions of $k=k_x+i k_y$. Such wavefunctions are not periodic under $k \to k+G$, but satisfy universal boundary conditions up to gauge redundancy and shifts in the unit cell center \cite{wang2021exact,Dong2023Many}. Classifying such $k$-space boundary conditions is mathematically intensive, hence why the proof of \eqref{eq:first_vortex_genform} and similar results are reserved for a separate manuscript. However, once the boundary conditions are determined, Liouville's theorem implies that two $k$-holomorphic periodic states, with the same boundary conditions and zeros, must be related by $\bk$-independent factors $\mathcal{N}_{0,1}(\br)$. We argue that this provides some intuition for how a general form such as \eqref{eq:first_vortex_genform} can emerge.

For readability, we recall the definition of first vortexability before showing that the wavefunctions \eqref{eq:first_vortex_genform} satisfy it. A band is first vortexable if 
\begin{enumerate}
\item[(P1)] ($\exists$ Partner Band) There exists an orthogonal vortexable band with projector $P_0$ 
(i.e. $P_0 z \Phi_0~=~z \Phi_0$ for all $P_0 \Phi_0 = \Phi_0$) such that
\begin{align}
\label{eq:defn_1vortexable_supp}
(P_1 + P_0) z\Phi_1 &= z \Phi_1
\end{align}
for all states $P_1\Phi_1 = \Phi_1$.
\item[(P2)] (Indecomposability) there is no alternative basis of wavefunctions $(\Phi_{0, \bk}', \Phi_{1, \bk}')$, where each band is vortexable, that also spans the two band subspace of $P_0 + P_1$.
\end{enumerate}

We now proceed to prove that wavefunctions of the form \eqref{eq:first_vortex_genform} are first vortexable, which will be our task for the remainder of this Appendix. It is essentially direct to show that the wavefunctions \eqref{eq:first_vortex_genform} obey (P1).
It suffices to show that $z \Phi_1$ is in the space spanned by wavefunctions of the form $\Phi_{0,1}$. 
Define 
\begin{equation}
    \tilde{\Phi}_1(\br)  =  \left(\overline{z}f(z) - 2 B_0^{-1} f'(z) \right) \mathcal{N}_{0}(\br) + f(z) \mathcal{N}_{1}(\br)
\end{equation}
so that $\Phi_1 = Q_0 \tilde{\Phi} = (1-P_0) \tilde{\Phi}$.

It again suffices to show that $z \tilde{\Phi}_1$ is in the space spanned by $\tilde{\Phi}_1$ and $\Phi_0$. Indeed, $\Phi_1$ and $\tilde{\Phi}_1$ only differ by the term $P_0 \tilde{\Phi}$ --- this term is in the vortexable partner band that maps to itself under $z$. We then have
\begin{equation}
\begin{aligned}
    z \tilde{\Phi}_1(\br) & = \left(\overline{z} (zf(z)) - 2 B_0^{-1} zf'(z) \right) \mathcal{N}_{0}(\br) + zf(z) \mathcal{N}_{1}(\br) \\
						  & = \underbrace{\left(\overline{z} g(z) - 2 B_0^{-1} g'(z) \right) \mathcal{N}_{0}(\br) + g(z) \mathcal{N}_{1}(\br)}_{\tilde{\Phi}_1[g(z)]} \\
                          &+ \underbrace{2 B_0^{-1} f(z) \mathcal{N}_0(\br)}_{\Phi_0[f(z)]},
\end{aligned}
\end{equation}
where $g(z) = z f(z)$ and $g'(z) = f(z) + z f'(z)$. The first two terms are of the form $\tilde{\Phi}_1$, while the last term is of the form $\Phi_0$. We therefore conclude that property (P1) is satisfied. 

Showing (P2) is more difficult, and we will need to move to momentum space and use holomorphicity in $k_x+ik_y$. We begin with a useful Lemma used repeatedly below. Here and below we use non-bolded symbols to refer to complex version of vectors, e.g. $k = k_x + i k_y, G = G_x+ i G_y$.
\lemma{A function of crystal momentum that satisfies $f(\bk + \bG) = f(\bk) - i \lambda \overline{G}$ under translation by reciprocal lattice vectors, for some $\mathbb{C} \ni \lambda \neq 0$, is not holomorphic in $k = k_x + i k_y$.}
\label{noholo}
\proof{Consider a contour integral of $f$ around the parallelogram boundary of the BZ, with side lengths $G_1$ and $G_2$, that vanishes for holomorphic functions by Cauchy's theorem. On the other hand, the shift-periodicity of $f$ relates the integrals on opposite sides of the parallelogram such that
\begin{equation}
    \oint_{\partial \bz} f(\bk) dk = -i \lambda(\overline{G_1} G_2 - \overline{G_2} G_1) =  \lambda A_\bz \neq 0.
    \label{eq:noholo_fromBCs}
\end{equation}
We conclude that $f$ is not holomorphic
\qed }

To use $k$-holomorphicity, we need explicit periodic Bloch states. Using the transformation rule \eqref{eq:translationofN} we see that $f_\bk(z) = e^{\frac{i}{2} \overline{k} z} \sigma(z + i B_0^{-1} k)$, where $\sigma$ is the (modified) Weierstauss $\sigma$-function, leads to Bloch periodic wavefunctions $\Phi_{0,1 \bk}(\br)$:
\begin{equation}
	\widehat{T}_{\v{R}} \Phi_{a \bk}(\br) = \Phi_{a \bk}(\br+\v{R}) = e^{i \bk \cdot \bR} \Phi_{a \bk}(\br), \quad a \in \{0,1\}.
\end{equation}
The corresponding periodic states $u_{a \bk}(\br) = e^{-i \bk \cdot \br}\Phi_{a \bk}(\br)$ are then given by

\begin{equation}
\begin{aligned}
    u_{0,k}(\br) & = f_k(\br)\mathcal{N}_0(\br)\\
    u_{1,\bk}(\br) & =  \left[\left(\ov{z} - 2 B_0^{-1} \partial_z -  \alpha(\bk) \right)f_k(\br)\right] \mathcal{N}_0(\br)  + f_k(\br) \mathcal{N}_1(\br)
    \end{aligned}
    \label{eq:periodicwfs_supp}
\end{equation}
where 
\begin{equation}
    f_k(\br) = e^{-\frac{i}{2} k \ov{z}} \sigma(z+i B_0^{-1} k).
\end{equation}
We have used that multiplication by $e^{i \bk \cdot \br}$ converts $e^{\frac{i}{2} \ov{k} z}$ to $e^{-\frac{i}{2} k \ov{z}} $.
The function $\alpha(\bk) = \braket{\Phi_{0\bk}|\tilde{\Phi}_{1,\bk}}$ comes from the action of the orthogonalization projector $Q_0$, and the functions $\mathcal{N}_{0,1}^l(\br)$ are functions of the layer and real-space but not $\bk$. 

The form of $f_k(\br)$ allows us to exchange real-space and momentum space derivatives: $\left(\ov{z} - 2 B_0^{-1} \partial_z \right)f_k(\br) = 2i\partial_k f_k(\br)$. 
We then obtain the $k$-holomorphic states
\begin{equation}
\begin{aligned}
    v_{0,k} = u_{0,k}, \,\, v_{1,k}(\br) = 2i\partial_k u_{0,k}(\br) + f_k(\br)  \mathcal{N}_1(\br)
    \end{aligned}
\end{equation}
by performing a linear combination to remove $\alpha(\bk)$. It is important to emphasize that the wavefunctions $v_{1,k}^{l}(\br)$ do not describe a vortexable band despite the fact that they are holomorphic in $k$. Indeed, they do not define a single Bloch band because $v_{1,k+G}$ is not proportional to $v_{1,k}$. Using
$\sigma(z+R)=\eta_{\bm{R}} e^{\frac{B}{2} \bar{R}(z+R/2)} \sigma(z)$, 
we have that 
\[
v_{0,k+G}(\br) = e^{-i \bG \cdot \br} \xi_\bG(k) v_{0,k}(\br)
\]
for $\xi_G(k) = \eta_\bG e^{\frac{1}{2} B^{-1} \ov{G}(k+\frac{G}{2}) }$. In contrast, 
\begin{equation}
\begin{aligned}
    v_{1,k+G}(\br) & = e^{-i \bG \cdot \br}\left( 2i \partial_k(\xi_{\bG}(k) v_{0,k} + \xi_{\bG}(k) f_k(\br) \mathcal{N}_1\right)  \\
    & = e^{-i \bG \cdot \br} \xi_{\bG}(k)( v_{1,k} + i B_0^{-1} \overline{G} v_{0,k}) 
    \end{aligned}
    \end{equation}
where we used $2i\xi_\bG^{-1} \partial_k \xi_\bG = i B_0^{-1} \overline{G}$. 
The existence of the above holomorphic gauge is an alternative proof of property (P1).

We must now prove (P2). Explicitly, gauge transforms in our present context have the form
\begin{equation}
\begin{pmatrix} w_{0,k} \\ w_{1,k} \end{pmatrix} = S_k^{-1} \begin{pmatrix} v_{0,k} \\ v_{1,k}\end{pmatrix}
\end{equation}
where $S_k$ are $k$-holomorphic, invertible matrics. We wish to show that there \textit{does not exist} any $S_k$ such that the $w_k$'s have diagonal boundary conditions
\begin{equation}
\begin{pmatrix} w_{0,k+G} \\ w_{1,k+G} \end{pmatrix} = e^{-i \bG \cdot \br} \begin{pmatrix} \tilde{\xi}_\bG^{(0)}(k) & 0 \\0 & \tilde{\xi}_\bG^{(1)}(k)   \end{pmatrix} \begin{pmatrix} w_{0,k} \\ w_{1,k}\end{pmatrix}.
\label{eq:diagbcs_tobecontr}
\end{equation}
Indeed, the existence of diagonal boundary conditions in momentum space in a holomorphic gauge is equivalent to the splitting of the two band subspace into two bands that are individually vortexable. 

Our strategy will be to assume diagonal boundary conditions \eqref{eq:diagbcs_tobecontr} and holomorphic $S_k$ and seek a contradiction. The boundary conditions of $v_{0,1k}$ that we computed above, phrased in matrix notation, are
\begin{equation}
    \begin{pmatrix} v_{0,k+G} \vspace{0.5em} \\ v_{1,k+G} \end{pmatrix} = e^{-i \bG \cdot \br}\xi_G(k) \begin{pmatrix} 1 & 0 \\ i B_0^{-1} \ov{G} & 1 \end{pmatrix} \begin{pmatrix} v_{0,k} \vspace{0.5em} \\ v_{1,k} \end{pmatrix}.
\end{equation}
The boundary conditions for $w_k$ are then related to the gauge transformation $S_k$ as 
\begin{equation}
    S_{k+G}\begin{pmatrix} \tilde{\xi}_\bG^{(0)}(k) & 0 \\0 & \tilde{\xi}_\bG^{(1)}(k)   \end{pmatrix} = 
    \xi_\bG(k)\begin{pmatrix} 1 & 0\\ i \ov{G} & 1 \end{pmatrix} S_k.
    \label{eq:attempt_diag}
\end{equation}

We first focus on the top left entry of \eqref{eq:attempt_diag}, which reads
\begin{equation}
S_{00}(k + G)\tilde{\xi}_{\bG}^{(0)}(k) = S_{00}(k) \xi_\bG(k),
\label{eq:00cpt}
\end{equation}
where $S_{00}$ is the top left entry of $S$.
Without loss of generality, assume that the Chern number $C_0$ of the band described by $w^{0,k}$ is zero or one (the Chern numbers must sum to two, and the Chern numbers of vortexable bands are positive). If $C_0=0$, then the band admits a tight-binding description with one delta-function-localized Wannier state per unit cell~\cite{ledwith2023vortexability}, and thus a gauge with $w_{0,k} = 1$, and $\tilde{\xi}^{(0)}_{\bG}(k)=1$. Since $S_k$ is invertible, we can demand that $h(k) =  \partial_k \log S_{00}(k)$ is holomorphic. However, by \eqref{eq:00cpt}, with $\tilde{\xi}^{(0)}_{\bG}(k)=1$, we have $h(k+G) = h(k) + \frac{1}{2} \ov{G}$, so that by Lemma \ref{noholo} $h(k)$ is not holomorphic. We therefore rule out the $C_0=0$ case, and focus on $C_0 = 1$. 

Vortexable bands with $C_0 =1$ have been classified~\cite{wang2021exact,Dong2023Many}, and have boundary conditions
$\tilde{\xi}_\bG^{(0)}(k) = \eta_\bG e^{\frac{1}{2} B^{-1} \ov{G}(k - k_0 + \frac{G}{2})} = \xi_\bG(k - k_0)$, for some offset $k_0$. Then, Lemma \ref{noholo} applied to $\log S_{00}(k)$ this time implies $k_0 = 0$, so that $S_{00}(k+G) = S_{00}(k)$ is periodic. Periodic holomorphic functions are constant, so we can set $S_{00}(k) = 1$ without loss of generality. The lower-left component of \eqref{eq:attempt_diag} then yields $S_{10}(k+G) = S_{10}(k) + i B_0^{-1} \ov{G}$, which again leads to a contradiction by Lemma \ref{noholo}. We have thus proven (P2). We conclude that the wavefunctions \eqref{eq:first_vortex_genform} used in the main text are first vortexable.

\section{Finding the Partner Band and determining first-vortexability}
\label{sec_app_partner}

This section explains how to algorithmically check if a band is first-vortexable for a given set of (perhaps numerical) band wavefunctions $\Phi_{1,\bk}(\br)$.
We proceed in three steps: (i) we describe how to find a candidate partner band given a conjectured first-vortexable band; (ii) we show that if the partner band constructed in the previous step is vortexable, then property (P1) is satisfied, i.e. the two-band space is vortexable; (iii) we explain how to find the instance of \eqref{eq:first_vortex_genform} that the band corresponds to, a procedure that will necessarily fail if the two-band vortexable subspace obtained in steps (i),(ii) is decomposable.

Step (i) comes from the observation that, from (P1), the partner band associated with a first vortexable band may be computed through $Q_1 z \Phi_1 = P_0 z \Phi_1 \propto \Phi_0$. We must therefore compute $Q_1 z \Phi_1$. Of course, acting with $z = x+iy$ is cumbersome numerically since typically one works with periodic boundary conditions and periodic wavefunctions $u_\bk(\br)$. To pass to these wavefunctions we observe that the product rule implies
\begin{equation}
    -i \bnabla_\bk \Phi_{1,\bk} = \br \Phi_{1,\bk} + e^{i \bk \cdot \br} (-i \bnabla_\bk) u_{1,\bk}.
\end{equation}
Acting with $Q_1$ annihilates the left hand side, which may be seen by writing the derivative as a finite difference between two nearby $\bk$-points.

We therefore have
\begin{equation}
    Q_1\br \Phi_{1,\bk} = Q_1 e^{i \bk \cdot \br} (i \bnabla_\bk) u_{1,\bk}.
\end{equation}
We now use that $Q_1 e^{i \bk \cdot \br} \nabla_\bk u_{1,\bk} = e^{i \bk \cdot \br} Q_1(\bk) \nabla_\bk \ket{u_\bk}$, where we used that $\bra{\Phi_{1,\bk'}} e^{i \bk \cdot \br} \nabla_\bk \ket{u_{1,\bk}} = 0$ for $\bk \neq \bk'$ by translation symmetry. Finally, adding the x component to $i$ times the $y$ component of the above vector equation and multiplying by $e^{-i\bk \cdot \br}$ yields
\begin{equation}
    u_{0,\bk}(\br) \propto e^{-i\bk \cdot \br} Q_1 z \Phi_{1,\bk} = 2i Q_1(\bk) \partial_k u_{1,\bk}(\br).
    \label{eq:findpartner}
\end{equation}

We pause to clarify notation in the above calculation; the projector $Q_1$ above is given by $Q_1 = 1 - \sum_\bk \ket{\Phi_{1,\bk}} \bra{\Phi_{1,\bk}}$, for normalized $\Phi_{1,\bk}$, which acts on states that satisfy Bloch periodicity. We also used corresponding projector $Q_1(\bk) = 1-\ket{u_{1,\bk}}\bra{u_{1,\bk}}$ that acts on the periodic wavefunctions $u_{1,\bk}$; while we use the same letter $Q_1$, the usage should be clear through the argument of $\bk$ and the examination of what the projector is acting on. 

We summarize step (i) by concluding that \eqref{eq:findpartner} suffices to compute a candidate partner band for a band that is conjectured to be first vortexable, in the sense that if the conjectured band is first vortexable then \eqref{eq:findpartner} yields the correct partner band. The reader may be concerned about computing the right hand side of \eqref{eq:findpartner} numerically given the derivative and the gauge redundancy of $u_{1,\bk}(\br)$. Indeed, even though \eqref{eq:findpartner} is gauge covariant, $u_{1 \bk} \to e^{i \theta_\bk} u_{1,\bk}$ induces $Q_1(\bk) \partial_k u_{1,\bk}(\br) \to e^{i \theta_\bk}Q_1(\bk) \partial_k u_{1,\bk}(\br) $, a sufficiently smooth gauge must be chosen in order to take the derivative. However, this gauge only needs to be smooth locally, and a locally smooth gauge in a neighborhood surrounding a point $\bk_0$ can be obtained through demanding that $\braket{u_{\bk_0}|u_\bk} > 0$.  

We now move on to using the partner band to check whether $\Phi_1$ is first vortexable, beginning with step (ii) and the associated property (P1). However, by the construction of step (i), the condition $(P_1 + P_0)z \Phi_1 = z \Phi_1$ is satisfied. It simply remains to check that the partner band constructed through \eqref{eq:findpartner} is vortexable. If it is, then the two band subspace spanned by $(\Phi_0,\Phi_1)$ is also vortexable.

Finally we must explain step (iii) and how check indecomposability, property (P2). For this, we use that the wavefunctions of first vortexable bands are given by \eqref{eq:first_vortex_genform}, and that zeroth and first vortexable pairs given by \eqref{eq:zero_vortex_genform},\eqref{eq:first_vortex_genform} must satisfy (P2). Thus, if we match our candidate wavefunctions to an instance of \eqref{eq:zero_vortex_genform},\eqref{eq:first_vortex_genform}, we have verified (P2). 

To obtain the trial wavefunction instance, we begin with the partner band. Note that below \eqref{eq:translationofN} we set $\br_0 \to 0$ for simplicity; here we must reinstate it in case the candidate parner band $\Phi_0$ is described by $\br_0 \neq 0$ in the computational choice of origin. However, we may calculate $\br_0$ by using that $\Phi_{0,\bk=0}(\br)$ should vanish at $\br = \br_0$ (and its translates by lattice vectors) due to \eqref{eq:zero_vortex_genform} and $\sigma(0)=0$. We then extract $\mathcal{N}_0(\br) = \Phi_{0,\bk=0}(\br)/\sigma(z - z_0)$, where $z_0$ is the complex coordinate corresponding to $\br_0$. This part of step (iii) is guarenteed to succeed since in step (ii) it was checked that $\Phi_0$ is vortexable. 

Next we must match the candidate $u_1$ to its general form \eqref{eq:periodicwfs_supp}; here we will find it slightly more convenient to match the periodic wavefunction. Note that there is a redundancy in $\mathcal{N}_1 \to \mathcal{N}_1 + \lambda \mathcal{N}_0$, since the second term drops under the projection operator $Q_0$ in \eqref{eq:first_vortex_genform}, or alternatively can be cancelled by $\alpha(\bk) \to \alpha(\bk) + \lambda$ in \eqref{eq:periodicwfs_supp}. Thus, $\mathcal{N}_1$ cannot be calculated directly, only $\mathcal{N}_1(\br) - \alpha(\bk)\mathcal{N}_0(\br)$ can. A further subtletly is that the formulae \eqref{eq:periodicwfs_supp} were stated without normalization and with gauge fixing, and distinct normalization and gauge fixing factors are required required to map onto the candidate states $u_0$ and $u_1$. Note that this issue did not enter for the extraction of $\mathcal{N}_0$, since we implicitly absorbed the normalization and gauge choice at $\bk=0$ into $\mathcal{N}_0$

We may get around these subtleties as follows. We will assume $\br_0 = 0$ again without loss of generality (we have already explained how to calculate it above if it is nonzero for some computational choice of origin). The candidate $u_1$, if it is first-vortexable, must have the form
\begin{equation}
\begin{aligned} 
    u_{1,\bk}(\br) =& \frac{1}{N_{1,\bk}} [(\overline{z} - 2 B_0^{-1} \partial_z)f_\bk(\br) \mathcal{N}_0(\br) \\
    &+ f_\bk(\br) (\mathcal{N}_1(\br) - \alpha(\bk) \mathcal{N}_0(\br)) ],
    \end{aligned}
    \label{eq:generalformu1k}
\end{equation}
which we wish to verify. We have an ansatz for $\mathcal{N}_0$, but $\mathcal{N}_{1,\bk} - \alpha(\bk) \mathcal{N}_0(\br)$ and the normalization $N_{1,\bk}$ are unknown. We will make use of the fact that the second group of terms vanishes when $\br = \br_k$, where $\br_k$ is such that the associated complex coordinate $z_k = -i B_0^{-1} k$, resulting in $f_k(\br_k)=0$, but the first group of terms generically does not. Thus, we may compute
\begin{equation}
    N_{1,\bk} = \frac{(\overline{z}_k - 2 B_0^{-1} \partial_z)f_\bk(\br)\big|_{\br = \br_k} \mathcal{N}^l_0(\br_k) }{u_{1,k}^l}
\end{equation}
for some fixed orbital index $l$. 

It is then straightforward to compute an expression
\begin{equation}
    X(\br,\bk) = \frac{N_{1,\bk} u_{1,\bk}(\br) - (\overline{z} - 2 B_0^{-1} \partial_z)f_\bk(\br) \mathcal{N}_0(\br)}{f_\bk(\br)}
\end{equation}
which should be of the form $\mathcal{N}_1(\br) - \alpha(\bk)\mathcal{N}_0(\br)$ if the band is first vortexable, according to \eqref{eq:generalformu1k}. $X$ has this form if and only if $\partial_{k_\mu} X \propto \mathcal{N}_0(\br)$, regarded as functions of $\br$. Indeed, if $\partial_{k_\mu} X \propto \mathcal{N}_0(\br)$ then Green's theorem applied to a line integral from $\bk_i$ to $\bk_f$ implies $X(\br,\bk_f) - X(\br,\bk_i) = (\alpha(\bk_f) - \alpha(\bk_i))\mathcal{N_0}(\br) $ for some function $\alpha(\bk)$. This determines $X$ up to a $\bk$-independent function of $\br$ that we may define as $\mathcal{N}_1(\br)$.  If the derivative of $X$ does not take this form, then the two band subspace can and should instead be written in terms of two zeroth-vortexable bands. 

\section{QGT of ``maximal" first vortexability}
\label{sec_app_max_QGT}
In this section, we compute the quantum geometric tensor of the combined space of a ``maximal" first vortexable band and its zeroth counterpart. By maximal, we specifically mean that there is no zeroth LL portion in the wavefunction other than what is required for the orthogonal basis. For example, in the language of the holomorphic periodic states of the previous section,
\begin{equation}
\begin{aligned}
    v_{k l}^{(0)} & = f_k(\br) \mathcal{N}_0(\br),  \\
    v_{k l}^{(1)} & = 2i \partial_k f_k(\br) \mathcal{N}_0(\br) + f_k(\br) \mathcal{N}_1(\br),
    \end{aligned}
\end{equation}
we demand $\mathcal{N}_1(\br) \to 0$. This corresponds to the limit $\gamma_1 \ell_B/v \to \infty$ in the Bernal graphene based Hamiltonians in the main text. In an orthonormal basis we then obtain
\begin{equation}
\begin{aligned}
    u_{0,\bk} & = \frac{1}{N_{0,\bk}}f_k(\br) \mathcal{N}_0(\br), \\
    u_{1,\bk} & =   \frac{1}{N_{1,\bk}} Q_0(\bk) \partial_k  u_{0,\bk}
    \end{aligned}
    \label{maximal_normalized_wfs_supp}
\end{equation}
where $Q_0(\bk) = 1 - \ket{u_{0,k}}\bra{u_{0,k}}$ orthogonalizes $u_{1,k}$ relative to $u_{0,k}$ and the factors $N_{m,\bk}$ are such that the wavefunctions are normalized. 

We recall property (V4), which states that vortexability of the bands is equivalent to $
\hat{\eta}^{\mu \nu}_{ab} = \frac{1}{2} \hat{\Omega}_{ab}(\delta_{\mu \nu} - i \varepsilon_{\mu \nu})$, where the Berry curvature is
\begin{equation}
    \hat{\Omega}_{ab}(\bk) = \bra{\partial_{k} u_{\bk b} } Q(\bk) \ket{\partial_{k} u_{\bk a}}
    \label{eq:nonAbBerry_supp}
\end{equation}
We now show that $\hat{\Omega}_{00} = \hat{\Omega}_{01} = \hat{\Omega}_{10} = 0$, in the maximal $\mathcal{N}^{(1)} \to 0$ limit, so that the only nonzero entry is $\hat{\Omega}_{11}$, as claimed in the main text. 

From \eqref{eq:nonAbBerry_supp}, it sufficies to show $Q(\bk) \ket{\partial_{k} u_{0,\bk}} = 0$. This in turn follows from writing
\begin{equation}
\begin{aligned}
    \ket{\partial_{k} u_{0,\bk}} & = P_0\ket{\partial_{k} u_{0,\bk}}+Q_0\ket{\partial_{k} u_{0,\bk}}, \\
    & = P_0\ket{\partial_{k} u_{0,\bk}}+N_{1,\bk}\ket{u_{1,\bk}},
    \end{aligned}
    \label{eq:split0thfirst}
\end{equation}
where we used \eqref{maximal_normalized_wfs_supp}. The first term is proportional to $u_{0,\bk}$ by construction while the second term is proportional to $u_{1,\bk}$, so that $Q(\bk)\ket{\partial_{k} u_{0,\bk}} = 0$ as claimed.

\section{Analytic Example at Zero Field}
\label{sec_app_analytical_example}
To demonstrate first vortexable bands without magnetic fields, we consider an analytically solvable (albeit artificial) example inspired by chiral twisted bilayer graphene. Let
\begin{equation}
\label{eq_TBG_1st_Ham}
H_{1}=\left(\begin{array}{cc}0 & D^{\dagger}_{1} \\ D_{1} & 0\end{array}\right), \quad D_{1}=\left(\begin{array}{cc} D_0 & \gamma_1 I_{2 \times 2} \\ 0 & D_0\end{array}\right).
\end{equation}
In particular, we assume that $D_0$ is the chiral Hamiltonian of twisted bilayer graphene (TBG), given by
\begin{equation}
\label{eq_TBG_Ham}
D_0 = \left(\begin{array}{cc} -2i \bar{\partial} & \alpha U_{\omega^*}(\bm{r}) \\ \alpha U_{\omega^*}(-\bm{r}) & -2i \bar{\partial}\end{array}\right), 
\quad U_{\omega}(\bm{r})=\alpha \sum_{j=0}^{2} \omega^{j} e^{-i \v{q}_j \cdot \bm{r}}
\end{equation}
with $\omega=e^{2 \pi i / 3}$ and $q_{j,x} + i q_{j,y} = -ie^{2\pi j /3} $. While \eqref{eq_TBG_1st_Ham} may appear strange due to the upper triangular hoppings connecting non-nearest-``layers", after a suitable unitary transformation it arises exactly as a $k_z=0$ band of three dimensional twisted AB-BA double bilayer graphene (see Sec~\ref{sec_app_3DATDBG}).
At each magic angle (such as $\alpha\approx 0.586$), $D_0$ has a zero mode~\cite{tarnopolskyorigin2019,ledwith2020fractional,wang2021chiral}
\begin{equation}
\label{intro_sigma_supp_bottompart}
\psi_{0,\bk}(\br)=\frac{\sigma\left(z+i B_0^{-1} k\right)}{\sigma(z)} e^{\frac{i}{2} \bar{k} z} \psi_{\Gamma}(\br),
\end{equation}
where $\sigma(z)$ is the Weierstrass $\sigma$-function~\cite{haldane2018modular,wang2021chiral}.
The Weierstrass sigma function $\sigma(z)$ satisfies $\sigma(-z)=-\sigma(z)$ and $\sigma(z+R)=\eta_{\bm{r}} e^{\frac{B}{2} \bar{R}(z+R / 2)} \sigma(z)$, where $\eta_{\bm{R}}=+1$ if $\bm{R} / 2$ is a lattice vector and $-1$ otherwise.

We can now construct $\Phi_{0,\bm{k}} = \big[ \psi_{0,\bm{k}},  0\big]^T$ and $\Phi_{1,\bm{k}} = \big[  \psi_{1,\bm{k}},  2i \gamma_1^{-1} \psi_{0,\bm{k}}\big]^T$, where
\begin{equation}
\psi_{1,\bk}(\bm{r})= \left[\bar{z}  - \frac{2B_0^{-1}\partial_{z} \sigma\left(z+i B_{0}^{-1} k\right)}{\sigma\left(z+i B_{0}^{-1} k\right) } - \alpha(\bk)\right]\psi_{0,\bk}(\bm{r}).
\end{equation}
with $\alpha(\bk)$ such that $\braket{\psi_{0,\bk} | \psi_{1, \bk} } = 0$.
Then the identity $D_0\psi_{1,\bk}=-\psi_{0,\bk}$ immediately implies $D_1 \Phi_{1,\bm{k}} = 0$, giving two bands of zero modes analogous to the LLL and 1LL.
We note $\psi_{1,\bk}$ obeys the Bloch condition because
the shift in $z \rightarrow z + R$ in the first term is cancelled out by
\begin{equation}
\partial_{z} \sigma(z+R)=-e^{\frac{B_{0}}{2} \bar{R}(z+R / 2)} \left[\partial_{z} \sigma(z)-\frac{B_{0}}{2} \bar{R} \sigma(z)\right].
\end{equation}

We numerically confirm that $\Phi_0$ and $\Phi_1$ are the zero-modes of $H_1$ [Eq.~\eqref{eq_TBG_1st_Ham}] at the magic angle of TBG.
There $H_1$ has four low-energy bands isolated from the remote bands, whose bandwidth vanishes at the same magic angle as TBG, $\alpha=0.586$ [Fig.~\ref{fig_Mk_TBG}(b)].
Due to the chiral symmetry of the Hamiltonian, the four eigenstates, denoted by $\Psi_{\pm E_{n}}$ with the energy $E_n > 0$ $(n=0,1)$, can be decomposed into sublattice polarized states such as $\Psi_{n\bm{k}}^{A,B}=\frac{1}{\sqrt{2}}\left(\Psi_{E_{n}, \bk} \pm \sigma_{z} \Psi_{E_{n}, \bk}\right)$.
In Fig.~\ref{fig_Mk_TBG}(b), we plot the bandwidth and the average of $1-\left|\braket{\Psi_{n\bk}^{A} | \Phi_{n\bk}}\right|$ over the BZ, denoted by $\langle T_{n}\rangle_{\rm BZ}$. At the magic angle $\Psi_{n\bm{k}}^{A} =\Phi_{n\bk}$ exactly, as expected.

\section{Three dimensional twisted AB-BA double bilayer}
\label{sec_app_3DATDBG}

 \begin{figure}
  \begin{center}
 \includegraphics[width=0.5 \textwidth]{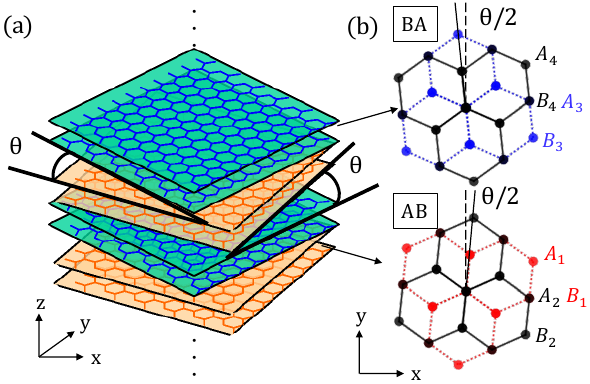}
    \caption{Schematic picture of three-dimensional alternatively ABBA twisted Bernal graphene.
    }
    \label{fig_3DATDBG_model}
  \end{center}
  \end{figure} 

The Hamiltonian [Eq.\eqref{eq_TBG_1st_Ham}] corresponds to the model where the layers are twisted by $(0,\theta,\theta,2\theta)$ and hosts a hopping between different sublattices of the next nearest layers.
Using the unitary transformation such as changing the order of layer basis $(|\sigma,1\rangle, |\sigma,2\rangle,|\sigma,3\rangle,|\sigma,4\rangle )$ into $(|\sigma,1\rangle, |\sigma,3\rangle,|\sigma,4\rangle,|\sigma,2\rangle )$, where $\sigma=\pm$ denotes $A/B$, we obtain the following Hamiltonian
\begin{equation}
\label{eq_3DATDBG_Ham}
\mathcal{H}_{\rm{3DATDBG}}=\mathcal{H}_{\rm{AB-BA}}+
\left(
\begin{array}{cccc}
0 & 0 & 0 & T^{\dagger}(\bm{r}) \\
0 & 0 & 0 & 0 \\
0 & 0 & 0 & 0 \\
T(\bm{r}) & 0 & 0 & 0
\end{array}\right).
\end{equation}
where the first term is the Hamiltonian of twisted AB-BA double bilayer (TDBG)~\cite{koshino2019band}
\begin{equation}
\mathcal{H}_{\rm{AB-BA}}=
\left(\begin{array}{cccc}
-i \boldsymbol{\sigma} \cdot \boldsymbol{\nabla} & \Gamma_1^\dagger & 0 & 0 \\
\Gamma_1 & -i \boldsymbol{\sigma} \cdot \boldsymbol{\nabla} & T^{\dagger}(\bm{r}) & 0 \\
0 & T(\bm{r}) & -i \boldsymbol{\sigma} \cdot \boldsymbol{\nabla} & \Gamma_1 \\
0 & 0 & \Gamma_1^\dagger & -i \boldsymbol{\sigma} \cdot \boldsymbol{\nabla}
\end{array}\right)
\end{equation}
with $\Gamma_1=(\sigma_x-i\sigma_y)\gamma_1/2$.
\begin{equation}
T(\bm{r})=\sum_{n=1}^{3} T_n e^{-i \bm{q}_{n} \cdot \bm{r}}
\end{equation}
where $T_{n+1}=\alpha\left(\sigma_{x} \cos n \phi+\sigma_{y} \sin n \phi\right)$ is the interlayer hopping in the chiral limit.
The second term is the interlayer coupling between layer 1 and 4, implying imposing the periodic boundary condition to the twisted AB-BA double bilayer in the stacking direction.
This model can be realized in three-dimensional alternatively twisted AB-BA double bilayer graphene shown in Fig.~\ref{fig_3DATDBG_model}.
In an infinite number of stackings of TDBG, the wavenumber $k_z$ along the stacking direction is a good quantum number to label the energy state.
At $k_z=0$, the Hamiltonian is equivalent to Eq.\eqref{eq_3DATDBG_Ham}.

\section{Trial wavefunction of zeroth and first vortexable state of strained Bernal graphene}
\label{sec_app_trial_wavefunction_strainedgraphene}

Let us construct the trial wavefunction of zeroth and first vortexable state of strained Bernal graphene.
At $\beta = 0$ the chiral symmetry of the model allows us to write
\begin{equation}
\begin{aligned}
    H_{\rm SBG} & = E_0 \begin{pmatrix} 0 & D_1^\dag \\ D_1 & 0 \end{pmatrix}, \,\,\, D_1 = \begin{pmatrix} D_0 & \gamma_1 \\ 0 & D_0 \end{pmatrix}  \\
    D_0 & = -2i \ov{\partial} + \alpha \A.
    \end{aligned}
\end{equation}
We will obtain the low energy subspace through sublattice polarized near-zero modes of $D_1$ and $D_1^\dag$. We will begin with $D_1$. At the $\Gamma$ point, there is an exact zero mode of $D_0$, $\psi_{0,\Gamma}(\br) = e^{-\alpha V(\br)}$, and an associated zero mode of $D_1$ (and thus $H$) by placing $\psi_{0,\Gamma}$ in the top layer component. We now create two near zero modes, for large $\alpha$, under the approximation $\psi_{0,\Gamma}(\br = 0) = e^{- 6 \alpha} \approx 0$; where here and below ``$\approx$" specifically means ``up to terms of $O\left(e^{-6 \alpha}\right)"$. The deviation of $O\left(e^{-6 \alpha}\right)"$ contributes to the total bandwidth of the four low energy bands.

We begin by constructing an approximate zero mode of $D_0$, following Ref. \cite{gao2023untwisting}. We will use the modified state $\psi^\eta = f_\eta(\br) e^{-  \alpha V(\br)}$, where $f_\eta(\br) = 0$ near $\br = 0$ but is $1$ otherwise (e.g. $f_\eta(\br) = \tanh \eta \abs{\br}$ for $\eta \gg 1$). Then, $D_0 \psi^\eta \approx 0$. Furthermore, since $\psi^\eta(\br = 0) = 0$ by construction, we have
\begin{equation}
    D_0 \mathcal{N}(\br) = 0, \qquad \mathcal{N}(\br) = \frac{\psi^\eta(\br)}{\sigma(z)}
\end{equation}
where $\sigma(z)$ is the Weierstrass $\sigma$-function~\cite{haldane2018modular,wang2021chiral,gao2023untwisting} that we used in Sec. \ref{sec_app_analytical_example}; see the discussion below \eqref{intro_sigma_supp_bottompart}. 
Since $\abs{\sigma(z)}^{-1}$ decreases exponentially as $\abs{z} \to \infty$, like the usual Gaussian factor of the LLL, we can now multiply by holomorphic functions, that necessarily blow up at infinity, and thereby construct a vortexable band $\psi_0 = f(z) \mathcal{N}(\br)$. 

These wavefunctions satisfy $D_0 \psi_0 \approx 0$ by construction. To construct Bloch states we write
\begin{equation}
\begin{aligned}
\psi_{0, \bk}(\br) & = f_\bk(z) \mathcal{N}_0(\br) \\
f_\bk(z) & = e^{\frac{i}{2} \overline{k} z} \sigma(z+i B_0^{-1} k).
\label{eq:approx_zero_D0_supp}
    \end{aligned}
\end{equation}

The construction \eqref{eq:approx_zero_D0_supp} directly yields an approximate zero mode of $D_1$, by placing $\psi_{0,\bk}(\br)$ in the top component, analogously to the zeroth LL of $B>0$ Bernal graphene and the analytical zero field model of the previous section. To obtain the second approximate zero mode, we write 
\begin{equation}
    \psi_{1, \bk}(\br) = Q_{\psi_0}(\overline{z} f_\bk(z) - 2B_0^{-1} \partial_z f_\bk(z))\mathcal{N}(\br),
\end{equation}
where $Q_{\psi_0}$ projects out the component proportional to $\psi_{0,\bk}$. Since $D_0 \psi_{0,\bk} \approx 0$ and $D_0 \psi_{1,\bk} \approx -\psi_{0,\bk}$, the states
\begin{equation}
\label{eq_strained_bernal_approxzero}
\Phi_{0,\bm{k}}(\bm{r}) =\left(\begin{array}{c}\psi_{0,\bm{k}}(\bm{r}) \\ 0\end{array}\right),
\Phi_{1,\bm{k}}(\bm{r})=\left(\begin{array}{c}  \psi_{1,\bm{k}}(\bm{r}) \\ 2i \gamma_1^{-1} \psi_{0,\bm{k}}(\bm{r})\end{array}\right),
\end{equation}
satisfy $D_1 \Phi_{(0,1),\bm{k}}(\bm{r}) \approx 0$, giving two almost flat bands of zero modes analogous to the LLL and 1LL. We may also write them in the general form we quoted for zeroth and first vortexable bands,
\begin{equation}
\begin{aligned}
    \Phi_{0,\bm{k}}(\bm{r}) & = f_{\bk}(z) \mathcal{N}_0(\br)  \\
    \Phi_{0,\bm{k}}(\bm{r}) & = Q_{\Phi_0}\left[ (\overline{z} f_\bk(z) - 2B_0^{-1} \partial_z f_\bk(z))\mathcal{N}_0(\br) + f_\bk(z) \mathcal{N}_1(\br) \right],
\end{aligned}
\end{equation}
where 
\begin{equation}
    \mathcal{N}_0(\br) = \mathcal{N}(\br)\begin{pmatrix} 1 \\ 0 \end{pmatrix}, \quad
    \mathcal{N}_1(\br)  =\mathcal{N}(\br) \begin{pmatrix} 0 \\ 2i \gamma_1^{-1} \end{pmatrix}.
\end{equation}

The two $B$-sublattice zero modes of $D_1^\dagger$ are topologically trivial. The exact zero mode of $D_0$, $\chi_{0,\Gamma} = e^{+\alpha V}$, is strongly localized at a single site $\br = \bR$ in each unit cell centered at the lattice vector $\bR$. Let us define $\chi_{0,\bR}(\br)$ to be equal to $\chi_{\Gamma}(\br)$ if $\br$ is in the unit cell centered at $\bR$, and zero otherwise. It is exponentially localized around $\br = \bR$, and satisfies $D_0 = \chi_{0,\bR}(\br) \approx 0$ as well; it should be thought of as a Wannier state of the strained monolayer. Now we construct
 \begin{equation}
     \mathcal{X}_{0,\bR}(\br) = \begin{pmatrix} \chi_{0,\bR}(\br) \\ 0 \end{pmatrix}, \quad \mathcal{X}_{1,\bR}(\br) = \begin{pmatrix} z \chi_{0,\bR}(\br) \\ 2i \gamma^{-1} \chi_{0,\bR}(\br) \end{pmatrix},
 \end{equation}
 as exponentially localized Wannier states at each site which satisfy $D_1 \mathcal{X}_{0,1 \bR}(\br) \approx 0$.

\section{ $\gamma_1$ dependence for $M(\mathbf{k})$}
\label{sec_app_Mk}
 \begin{figure}
  \begin{center}
 \includegraphics[width=0.5 \textwidth]{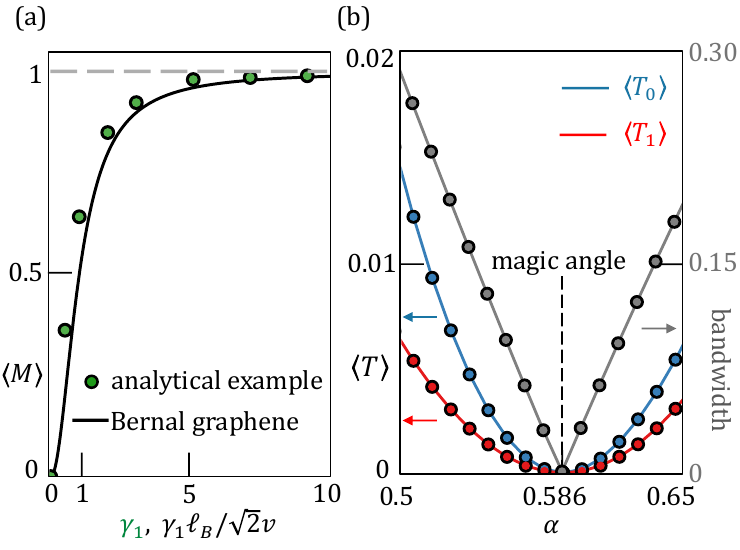}
    \caption{(a)$\langle M \rangle_{\rm BZ}$ for the analytical example [Eq.\eqref{eq_TBG_1st_Ham}] (green points) and Bernal graphene with magnetic field [Eq.\eqref{eq_Mk_BG}] (black line) as a function of $\gamma_1$, and $\gamma_1 \ell_B/\sqrt{2}v$, respectively. (b) The gray line represents the bandwidth of the set of four low-energy bands of Eq.\eqref{eq_TBG_1st_Ham} as a function of $\alpha$ for $\gamma_1=1$. The bandwidth becomes zero at the magic angle of TBG, $\alpha=0.586$.
The blue and red lines are the averaged overlaps $\langle T_{n}\rangle_{\rm BZ}$ for $n=0$ and $n=1$, respectively.
    }
    \label{fig_Mk_TBG}
  \end{center}
  \end{figure} 

In this section, we compute the
``maximality index" which measures how close one is to the ``maximal" first~LL:
\begin{equation}
    M(\bk) = \frac{\abs{\lambda_1(\bk) - \lambda_2(\bk)}}{\lambda_1(\bk) + \lambda_2(\bk)},
\end{equation}
where $\lambda_{1,2}$ are the eigenvalues of the $2 \times 2$ non-Abelian Berry curvature. The previous subsection implies $M(\bk) = 1$ when $\mathcal{N}_1 \to 0$. Here we will compute $M$ for Bernal graphene with constant $B>0$ analytically. We will also discuss the $M(\bk)$ of the chiral TBG example, which ends up closely related. 

Bernal graphene in an external magnetic field has a continuous magnetic translation symmetry; the magnetic translation operators are $T(\bm{l}) = e^{-i \xi_{\bm{l}}(\bm{r})}t(\bm{l})$ where $t(\bm{l})=e^{i\bm{l} \cdot \boldsymbol{\nabla}}$ is the ordinary, $B=0$ translation operator. The phase satisfies $\bm{A}(\bm{r}+\bm{l})=\bm{A}(\bm{r})+\nabla \xi_{\bm{l}}(\bm{r})$, where $\v{A}$ is the vector potential. The magnetic translation operators satisfy
\begin{equation}
    T(\v l_1) T(\v l_2) = T(\v l_2) T(\v l_1) e^{i B \hat{z} \cdot \v l_1 \times \v l_2}
    \label{eq:magneticalgebra}
\end{equation}
A commuting subset, generated by $T_{\bR_1}$, $T_{\bR_2}$ with $B \hat{z} \cdot \bR_1 \times \bR_2 = 2\pi$, can then be used to define Bloch states $\Phi_{\bk}(\br)$. The algebra \eqref{eq:magneticalgebra} implies that magnetic translations at other $\ell \neq \bR$ act as ``ladder operators" on the Bloch states, $T_{\v l} \Phi_{\bk} \propto \Phi_{\bk - B \hat{z} \times \v l}$ (see e.g. Appendix B of Ref. \cite{Dong2023Many} for a pedagogical discussion). Thus, all Bloch momenta are symmetry related. We will shortly use this fact to specialize to calculating the Berry curvature at $\bk = 0$.

The periodic states $\mathcal{U}_{m,\bk} = e^{-i \bk \cdot \br} \Phi_{m,\bk}(\br)$ of Bernal graphene have the form
\begin{equation}
\begin{aligned}
\mathcal{U}_{0,\bk}& =\left(\begin{array}{c}u_{0,\bk} \\ 0\end{array}\right), \\
\mathcal{U}_{1,\bk}& =\frac{1}{\sqrt{\gamma_1^2 + 2 (v \ell_B^{-1})^2}}\left(\begin{array}{c} \gamma_1 u_{1,\bk} \\ \sqrt{2} v \ell_B^{-1} u_{0,\bk}\end{array}\right),
\end{aligned}
\end{equation}
where $u_{0, \bk}$ is the periodic state of the usual LLL and $u_{1,\bk}$ is that of the usual first LL, obtained from the lowest through the raising operator. Both states are normalized.
We will shortly use 
\begin{equation}
    u_{1,\bk} = \sqrt{B} Q_0(\bk) \partial_k u_{0,\bk}
    \label{eq:u1normalized}
\end{equation}
which can be computed directly or indirectly. The indirect route begins by writing $u_{1,\bk} = \lambda Q_0(\bk) \partial_k u_{0k}$. One then solves for $\lambda = \sqrt{B}$ through $1 = \braket{u_{1,\bk}|u_{1,\bk}} = \abs{\lambda}^2 \bra{\partial_k u_{0,\bk}}Q(\bk) \ket{\partial_k u_{0,\bk}} = \abs{\lambda}^2 \Omega_{\text{LLL}} $, where we have identified $\Omega_{\text{LLL}} = 1/B$ as the Berry curvature of the LLL. 
We will also use
\begin{equation}
\begin{aligned}
\langle u_{1,\bk}|\partial_k u_{0,\bk} \rangle & = \sqrt{B} \langle \partial_k u_{0,\bk}|Q_0(\bk)|\partial_k u_{0,\bk} \rangle, \\
& = \sqrt{B}\Omega_{\text{LLL}} = 1/\sqrt{B}.
\end{aligned}
\end{equation}

We are now ready to compute the non-Abelian Berry curvature
\begin{equation}
\hat{\Omega}_{mn}(\bk)=\left\langle\partial_{k} \mathcal{U}_{n,\bk}|Q(\bk)| \partial_{k} \mathcal{U}_{m,\bk}\right\rangle.
\end{equation}
Let us fix $\bk = 0$ without loss of generality, so that we can more easily apply rotation symmetry (recall that all momenta are related by magnetic translation symmetry, such that $\hat{\Omega}$ is $\bk$-independent). We note that $\mathcal{U}_{1,\bk=0}$ has a different angular momentum than $\mathcal{U}_{0,\bk=0}$. This implies that the off-diagonal elements of $\hat{\Omega}$ must vanish.

The diagonal terms follow from 
\begin{equation}
\begin{aligned}
\hat{\Omega}_{00}& = \langle \partial_k\mathcal{U}_{0,\bk}|(1-P_0-P_1)|\partial_k \mathcal{U}_{0,\bk} \rangle\\
& = \langle \partial_k \mathcal{U}_{0,\bk}|Q_0|\partial_k \mathcal{U}_{0,\bk} \rangle - \langle \partial_k \mathcal{U}_{0,\bk} |P_1|\partial_k \mathcal{U}_{0,\bk} \rangle\\
& = \Omega_{\text{LLL}} - \frac{\gamma_1^2}{\gamma_1^2 + 2  (v \ell_B^{-1})^2}\abs{\langle u_{1,k}|\partial_k u_{0,k} \rangle}^2 \\
& =B^{-1}\frac{2(v \ell_B^{-1})^2}{\gamma_1^2 + 2 (v \ell_B^{-1})^2}.
\end{aligned}
\end{equation}

Because the trace of the non-Abelian Berry curvature must be $2/B$, to be consistent with $C=2$ for the two-band system, we must have
\begin{equation}
\hat{\Omega}_{11}=B^{-1}\frac{2 (v \ell_B^{-1})^2+2\gamma_1^2}{\gamma_1^2 + 2 (v \ell_B^{-1})^2}.
\end{equation}

We therefore obtain
\begin{equation}
\label{eq_Mk_BG}
M = \frac{\gamma_1^2}{\gamma_1^2 + 2 (v \ell_B^{-1})^2},
\end{equation}
as claimed in the main text.
At $\gamma_1 \rightarrow 0$ which corresponds to the decoupling of Bernal graphene into two identical monolayer graphene, the ingredients of non-Abelian Berry curvature are $\hat{\Omega}_{01}=\hat{\Omega}_{10}=0$ and $\hat{\Omega}_{00}=\hat{\Omega}_{11}=1/B$ so the maximal index is $M=0$.
On the other hand, at $\gamma_1 \rightarrow \infty$,  $\hat{\Omega}_{00}=\hat{\Omega}_{01}=\hat{\Omega}_{10}=0$ and $\hat{\Omega}_{11}=2/B$, leading to $M=1$.

We numerically calculate the $\langle M(\bm{k}) \rangle_{\rm BZ}$ for the analytical example [Eq.\eqref{eq_TBG_1st_Ham}] at magic angle $\alpha=0.586$ in Fig.\ref{fig_Mk_TBG}.
The index depends on $\gamma_1$ almost as same as that of Bernal graphene $B>0$.

\section{Self-consistent Hatree-Fock calculation}
\label{sec_app_SCHF}
We consider realistic gate-screened Coulomb interactions:
\begin{equation}
\hat{H}_{\rm tot}=\hat{H}_{SBG}+\frac{1}{2 A} \sum_{\boldsymbol{q}} V_{\boldsymbol{q}}: \hat{\rho}_{\boldsymbol{q}} \hat{\rho}_{-\boldsymbol{q}}:, \quad V_{q}=\frac{2 \pi \tanh (q d)}{\epsilon_{r} \epsilon_{0} q}
\end{equation}
with density operator $\rho_{\bm{q}}$ at wavevector $\bm{q}$, normal ordering relative to filling $\nu = 0$, sample area $A$, gate distance $d=\SI{250}{\angstrom}$, and relative permittivity $\epsilon_r = 15$. To obtain the two isolated Chern bands, we employ self-consistent Hartree-Fock (SCHF) calculations on $24 \times 24$ unit cells. We assume spin and valley polarization~\cite{gao2023untwisting}, and project to well-isolated subspace of the three highest valence bands.

\bibliography{flat_1stLL.bib}

\end{document}